\documentclass[nohyperref]{article}
\PassOptionsToPackage{numbers, compress}{natbib}
\usepackage{iclr2025_conference, times} 

\iclrfinalcopy

\usepackage[utf8]{inputenc} %

\usepackage{graphicx}

\usepackage[T1]{fontenc}

\usepackage[hidelinks,colorlinks]{hyperref}       %
\usepackage{url}            %
\usepackage{booktabs}       %
\usepackage{amsfonts}       %
\usepackage{nicefrac}       %
\usepackage{microtype}      %

\IfFileExists{headers/config/showoverfull.config}{
	\overfullrule=1cm
}{
}

\usepackage{marginnote}

\usepackage[backgroundcolor=none,linecolor=red,textsize=footnotesize]{todonotes}

\usepackage{etoolbox}

\newbool{includeappendix}
\setbool{includeappendix}{true} %
\IfFileExists{headers/config/noappendix.config}{
	\setbool{includeappendix}{false}
}{}

\newif\ifincludeappendixx
\ifbool{includeappendix}{
	\includeappendixxtrue
}{
	\includeappendixxfalse
}

\usepackage{xr} %
\usepackage{filecontents}

\ifbool{includeappendix}{}{
	\input{appendix-labels-loader}

	\externaldocument{appendix-labels}
}

\newcommand{\eg}{e.g., }
\newcommand{\ie}{i.e., }

\usepackage{acro} %

\usepackage{listings}

\usepackage{textcomp}

\usepackage{xcolor}

\usepackage[scaled=0.8]{beramono}

\definecolor{ckeyword}{HTML}{7F0055}
\definecolor{ccomment}{HTML}{3F7F5F}
\definecolor{cstring}{HTML}{2A0099}

\lstdefinestyle{numbers}{
	numbers=left,
	framexleftmargin=20pt,
	numberstyle=\tiny,
	firstnumber=auto,
	numbersep=1em,
	xleftmargin=2em
}

\lstdefinestyle{layout}{
	frame=none,
	captionpos=b,
}

\lstdefinestyle{comment-style}{
	morecomment=[l]//,
	morecomment=[s]{/*}{*/},
	commentstyle={\color{ccomment}\itshape},
}

\lstdefinestyle{string-style}{
	morestring=[b]",%
	morestring=[b]',%
	stringstyle={\color{cstring}},
	showstringspaces=false,%
}

\lstdefinestyle{keyword-style}{
	keywordstyle={\ttfamily\bfseries},
	morekeywords={
		function,
		constructor,
		int,
		bool,
		return,
		returns,
		uint
	},
	morekeywords = [2]{},
	keywordstyle = [2]{\text},
	sensitive=true,
}

\lstdefinestyle{input-encoding}{
	inputencoding=utf8,
	extendedchars=true,
	literate=
	{ℝ}{$\reals$}1%
	{→}{$\rightarrow$}1%
	{α}{$\alpha$}1%
	{β}{$\beta$}1%
	{λ}{$\lambda$}1%
	{θ}{$\theta$}1%
	{ϕ}{$\phi$}1%
}

\lstdefinestyle{escaping}{
	moredelim={**[is][\color{blue}]{\%}{\%}},
	escapechar=|,
	mathescape=true
}

\lstdefinestyle{default-style}{
	basicstyle=\fontencoding{T1}\ttfamily\footnotesize,
	style=numbers,
	style=layout,
	style=comment-style,
	style=string-style,
	style=keyword-style,
	style=input-encoding,
	style=escaping,
	tabsize=2,
	upquote=true
}

\lstdefinelanguage{BASIC}{
	language=C++,
	style=default-style
}[keywords,comments,strings]%

\usepackage[utf8]{inputenc} %
\usepackage[T1]{fontenc}    %
\usepackage{algcompatible}
\usepackage{caption}
\usepackage{algpseudocode}
\usepackage{enumerate}
\usepackage{url}            %
\usepackage{booktabs}       %
\usepackage{amsfonts}       %
\usepackage{nicefrac}       %
\usepackage{microtype}      %
\usepackage{xcolor}         %
\usepackage[ruled,vlined]{algorithm2e}
\usepackage{tikz,booktabs,multirow,enumitem,bm}
\usepackage{colortbl}
\usepackage{tabularx}
\usepackage{subcaption}
\usepackage{wrapfig}
\usepackage{tabulary}
\usepackage{amsmath,amsthm,amsfonts, bbm}
\usepackage{makecell}

\usepackage{siunitx}

\newcommand*\numcir[1]{\raisebox{.5pt}{\textcircled{\raisebox{-.9pt} {#1}}}}

\usepackage{amsmath,amsfonts,bm}
\usepackage{amsthm}

\def\1{\bm{1}}

\DeclareMathAlphabet{\mathsfit}{\encodingdefault}{\sfdefault}{m}{sl}
\SetMathAlphabet{\mathsfit}{bold}{\encodingdefault}{\sfdefault}{bx}{n}

\newcommand{\KGW}{\textsc{KGW-D}\xspace}
\newcommand{\KTH}{\textsc{KTH-D}\xspace}
\newcommand{\unremovable}{\textsc{Unremovable}\xspace}
\newcommand{\gaussmark}{\textsc{GaussMark}\xspace}
\newcommand{\llama}{\textsc{Llama2-7B}\xspace}

\makeatletter
\newcommand*{\centernot}{%
  \mathpalette\@centernot
}
\def\@centernot#1#2{%
  \mathrel{%
    \rlap{%
      \settowidth\dimen@{$\m@th#1{#2}$}%
      \kern.5\dimen@
      \settowidth\dimen@{$\m@th#1=$}%
      \kern-.5\dimen@
      $\m@th#1\not$%
    }%
    {#2}%
  }%
}
\makeatother

\newcolumntype{x}[2]{S[table-format=#1.#2,table-auto-round]}

\newcolumntype{y}[2]{>{\small} S[table-format=#1.#2,table-auto-round]}

\definecolor{hyperlinkblue}{HTML}{0000AA}
\hypersetup{citecolor=hyperlinkblue} %

\usepackage[capitalize,noabbrev]{cleveref}

\makeatletter
\AddToHook{cmd/appendix/before}{\crefalias{section}{appendix}}
\AddToHook{cmd/appendix/before}{\crefalias{subsection}{appendix}}
\AddToHook{cmd/appendix/before}{\crefalias{subsubsection}{appendix}}
\makeatother

\crefformat{equation}{Eq.~(#2#1#3)}

\crefformat{section}{\S#2#1#3}

\crefrangeformat{section}{\S#3#1#4\crefrangeconjunction\S#5#2#6}

\crefmultiformat{section}{\S#2#1#3}{\crefpairconjunction\S#2#1#3}{\crefmiddleconjunction\S#2#1#3}{\creflastconjunction\S#2#1#3}

\newcommand{\crefrangeconjunction}{--}

\crefname{listing}{Lst.}{listings}
\crefname{line}{Lin.}{Lin.}
\crefname{appendix}{App.}{App.}

\newcommand{\appref}[1]{%
	\ifbool{includeappendix}{\cref{#1}}{the appendix}%
}
\newcommand{\Appref}[1]{%
	\ifbool{includeappendix}{\cref{#1}}{The appendix}%
}

\title{{Towards Watermarking of Open-Source LLMs}}

\author{%
  Thibaud Gloaguen, Nikola Jovanović, Robin Staab, Martin Vechev\\
  ETH Zurich\\
  Correspondence to: \texttt{tgloaguen@student.ethz.ch}, \texttt{nikola.jovanovic@inf.ethz.ch}
}

\begin{document}

\sisetup{
text-series-to-math = true,
}

\maketitle

\begin{abstract}
\vspace{-0.1in}
While watermarks for closed LLMs have matured and have been included in large-scale deployments, these methods are not applicable to open-source models, which allow users full control over the decoding process. 
This setting is understudied yet critical, given the rising performance of open-source models.
In this work, we lay the foundation for systematic study of open-source LLM watermarking.
For the first time, we explicitly formulate key requirements, including \emph{durability} against common model modifications such as model merging, quantization, or finetuning, and propose a concrete evaluation setup. 
Given the prevalence of these modifications, durability is crucial for an open-source watermark to be effective.
We survey and evaluate existing methods, showing that they are not durable.  
We also discuss potential ways to improve their durability and highlight remaining challenges.  
We hope our work enables future progress on this important problem. 

\end{abstract} 

\section{Introduction}
\label{sec:intro}

As highlighted by recent AI regulations~\citep{aia}, watermarking of large language models (LLMs) to track their outputs is an increasingly important research topic.
Building on earlier work~\citep{aar,kgw,stanford}, recently proposed methods~\citep{synthid-text} have been deployed in large-scale production systems~\citep{synthid}, showing that LLM watermarking is reaching maturity. 
Most existing methods are based on modifying the decoding procedure of the LLM to imprint a later detectable watermark signal. 
As such, these \emph{generation-time watermark mechanisms} are designed for closed models served via an API.  

\paragraph{Open-source LLM watermarking}
This makes most advancements in LLM watermarking fundamentally inapplicable to open-source models (\emph{OSM}). Such models allow users white-box access, including full control over the decoding procedure, which they can use to disable any generation-time watermarking mechanism. 
This is becoming increasingly important as the gap between closed and open models is narrowing---latest versions of Llama~\citep{llama3}, Qwen~\citep{qwen}, and DeepSeek~\citep{deepseek} models have nearly surpassed the performance of best closed-source models.
If this trend continues, malicious parties will be able to circumvent any watermarked API by using OSM to generate high-quality unwatermarked text.
This makes \textit{open-source model watermarking}, i.e., the question of how model providers can embed watermarks directly into their open-weight models, a key focus for the watermarking community.

\paragraph{Prior work}
While OSM watermarking has been recognized as one of the most critical problems in GenAI security~\citep{survey2,genaisurvey,wmsok,whatliesahead}, it has not been systematically studied. 
While there have been a few attempts to embed the watermark directly into model weights~\citep{unremovable, learnability, rlwatermark, gaussmark}, in many works, the specific challenges of OSM watermarking are only a secondary focus.  
Even when OSM is the main focus, there is a lack of clarity regarding problem formulation and evaluation targets---some works consider random adversaries~\citep{unremovable}, while others focus only on the finetuning of the watermarked model, restricted to broad-domain data~\citep{learnability}.

\begin{figure}[t]
    \centering
    \includegraphics[width=\textwidth]{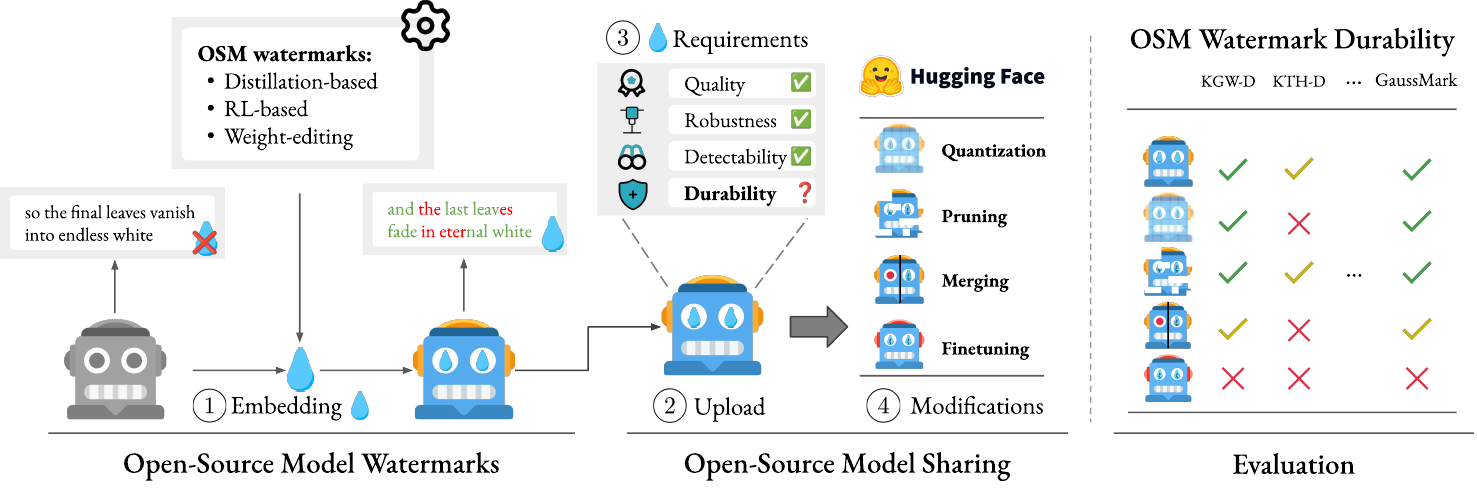}
    \caption{Definition and evaluation of OSM watermark durability. 
    \numcir{1}: Given a base unwatermarked model, a watermark is embedded into its weights. 
    \numcir{2}: The model is uploaded to a model-sharing platform like Hugging Face.
    \numcir{3}: The model is tested against the established requirements for generation-time watermarks. 
    \numcir{4}: Yet, third-party users modify the weights of the model through quantization, pruning, merging, and finetuning, and may distribute the modified model. 
    We ask: \emph{are such modified models still watermarked}? 
    To evaluate this, we introduce a new requirement: {durability}, and propose a systematic evaluation procedure based on the most common model modifications.
    }
    \label{fig:accept}
    \vspace{-0.1in}
\end{figure}

\paragraph{This work}
In this work, we aim to provide the first systematic study of the problem of open-source LLM watermarking, laying the foundation for future research.
\cref{fig:accept} illustrates the OSM watermarking setting and our contributions.
First, we revisit the requirements for generation-time LLM watermarks and discuss how they apply to the OSM case.
Aiming to concretize the specific challenges of the OSM setting, we define a new requirement: \emph{durability against common model modifications}.
While this was not a concern for closed models, considering such non-adversarial changes is crucial for OSM, as these models are typically finetuned, quantized or modified in other ways (\cref{sec:durability}).
As the key goal of watermarking is to protect \emph{every} model output, a watermark that is not durable against such changes fails to fulfill its purpose in most realistic scenarios.
 
We complement this definition by proposing an evaluation procedure for OSM watermark durability based on a collection of the most common model modifications. In our evaluation of all current OSM watermarks (and a new variant that we propose in~\cref{sec:eval}), we find that while most methods endure some modifications, no method is truly durable---establishing durability is a challenging but worthwhile requirement for OSM watermarks.
To motivate future work in this direction, we propose a proof-of-concept experiment on a \textsc{GPT-2} architecture to explore ways to improve watermark durability.  
Our work is a first step toward more systematic study of OSM watermarks.

\paragraph{Key contributions}
Our main contributions are: 
\begin{itemize}
    \item We reinterpret LLM watermark requirements in the context of open-source models, and introduce the critical requirement of watermark durability~(\cref{sec:requirements,sec:durability}).
    \item We survey current OSM watermarks, propose a new variant, and systematically evaluate all methods for durability, concluding that no watermark is truly durable~(\cref{sec:methods,sec:eval}).
    \item We present a proof-of-concept experiment that explores ways to improve watermark durability and identify challenges that remain, motivating future work~(\cref{sec:wm_from_scratch}). 
\end{itemize}

\section{Requirements for Watermarking of Open-source Models}
\label{sec:requirements}

We first recall generation-time watermarks and their requirements (\cref{ssec:requirements_post_hoc}) and detail the challenges faced by OSM watermarks, along with the new requirements induced by those challenges (\cref{ssec:osm_watermark_requirements}).

\paragraph{Notation}
We define an LLM as an autoregressive model over a vocabulary $\Sigma$, parameterized by its weights $\theta \in \Theta$ where usually $\Theta \subset \mathbb{R}^d$.
Given a text $x \in \Sigma^*$, with $\Sigma^*$ the Kleene closure of $\Sigma$, the next token distribution according to the LLM is given by $p_{\theta}(.|x_{<t})$. 
We also refer to $y(x, \theta)$ as a sampled completion from the model $\theta$ for the prompt $x$.
Lastly, we note $p_{\theta}(x):= \prod_{t=1}^{|x|} p_{\theta}(x_t|x_{<t})$.

\subsection{Established Requirements for Generation-time Watermarks}
\label{ssec:requirements_post_hoc}

Next, we introduce generation-time watermarking methods and recall the requirements that have been established in prior work.

\paragraph{Generation-time watermarks} 
A generation-time watermark $w$ is defined by a triple $(f_w, \xi_w, \mathcal{D}_w)$. 
$\xi_w \in \mathbb{N}$ is a private key (used to seed a pseudo-random function).
$f_w$ is a mapping from $\Delta(\Sigma) \times \Sigma^* \times \mathbb{N}$ to $\Delta(\Sigma)$ that takes a next-token probability distribution, a sequence of tokens, and the private key, and returns a watermarked next-token probability distribution.
Lastly, $\mathcal{D}_w: \Sigma^* \times \mathbb{N} \rightarrow \{0,1\}$ is a \emph{watermark detector} that, given a text $x$ and the private key $\xi_w$, returns $1$ if $x$ is watermarked and $0$ otherwise. 
Hence, given a model $\theta$ and a text $x$, to generate watermarked text, at each step $t$ of the generation process, instead of sampling the next token from $p_{\theta}(.|x_{<t})$ we sample the next token from $f_w(p_{\theta}(.|x_{<t}), x_{<t}, \xi_w)$.
We may omit the dependency on $x_{<t}$ in $f_w$ by writing $f_w(p_{\theta}(.|x_{<t}), \xi_w)$.

\paragraph{Requirements} 
On top of watermark strength, \ie
the ability of the watermarking algorithm to produce text in which the watermark is detectable, four key requirements of watermarks have been identified in previous works \citep{stanford,unbiased,dipmark}.
\begin{enumerate}
    \item \emph{Quality}: A watermark should not significantly degrade the quality of the model outputs.  
    A proxy for quality in the literature is distortion-freeness \citep{stanford,unbiased,orzamir,dathathri2024scalable}.  
    In expectation over the private key $\xi$, the next-token distribution (or, in the stronger case, the next-sequence-of-tokens distribution) should be the same between the original model and the watermarked model.
    \item \emph{Robustness} \citep{kgw2}: Given a watermarked text $x$, robustness measures how edits to the text (\eg token or word insertion, deletion, substitution, and paraphrasing) affect the accuracy of the watermark detector.
    \item \emph{Undetectability} \citep{orzamir,crafted_prompt,detection}: Measures the ability of third-party actors to detect the presence of the watermark without the private key.  
    \item \emph{Security}: This encompasses the vulnerability of the watermarking algorithm to spoofing and stealing attacks \citep{watermark_stealing,bileve,strengths}, where a third party tries to impersonate the watermark without having the private key $\xi_w$.
\end{enumerate}

\subsection{How does the Open-source Setting Differ?}
\label{ssec:osm_watermark_requirements}

With an open-source model $\theta$, the end user has direct access to $p_{\theta}(.|x_{<t})$.
Hence, a malicious user can choose to sample the next token directly according to $p_{\theta}(.|x_{<t})$ rather than $f_w(p_{\theta}(.|x_{<t}), \xi)$, thereby producing non-watermarked text.
Thus, for an open-source model, the watermark must be directly embedded into the model's weights (\ie $p_{\theta}(.|x_{<t})$). 
This conceptual difference from generation-time watermarks requires an updated and newly interpreted set of requirements.

\paragraph{OSM watermarks}
We define a watermark for open-source models $w$ as a triple $(g_w, \xi_w, \mathcal{D}_w)$, where $\xi_w \in \mathbb{N}$ is a private key, $g_w: \Theta \times \mathbb{N} \rightarrow \Theta$ is a mapping from the initial model to its watermarked version, and $\mathcal{D}_w$ is the watermark detector.
This means that the watermark is now embedded into the weights of the model and sampling text according to $p_{\theta}(.|x_{<t})$ generates watermarked text.

\paragraph{Requirements for OSM watermarks}
Based on the requirements for generation-time watermarks, previous works have already identified \textit{Quality} and \textit{Robustness} as requirements that directly extend to OSM watermarks.  
\textit{Undetectability} has not been addressed in prior works but similarly remains an important concern. 
In particular, distillation-based OSM watermarks (see \cref{sec:methods}) inherit the detectability issues of the generation-time watermark they are distilling \citep{detection,crafted_prompt}.
The other current OSM watermarks that we introduce in \cref{sec:methods} either have similar issues or, while no detection methods on them have been demonstrated, also do not offer guarantees of undetectability.

In contrast to generation-time watermarks, \textit{Security}, in particular spoofing, is not considered a fundamental issue for OSM watermarks.
For standard LLM watermarks, spoofing undermines the model provider's credibility, allowing attackers to produce malicious texts falsely attributed to them. As the user has direct control over the OSM watermarked model, more powerful and easier jailbreaking attacks \citep{prefill_jailbreak} directly enable the generation of harmful watermarked text from such models. 
Independently, spoofing threatens the integrity of multi-bit watermarks \citep{multibit1,multibit2,multibit3} by enabling impersonation---this is not applicable to OSMs, where generally only a single fixed model is released. 

\paragraph{OSM watermark durability}
The watermarked model being open-source introduces an additional key requirement: \emph{durability} against model modifications (see \cref{sec:durability}).
Durability measures how edits to the model parameters affect watermark detection.
As discussed in \cref{sec:intro}, no generation-time watermark is durable in this way, as it can be easily removed by changing the sampling algorithm.
In prior works, durability has either been defined in unrealistic scenarios \citep{unremovable}, tested against incomplete adversaries \citep{learnability}, or not considered at all.
Yet, we argue that a thorough consideration of durability is crucial for OSM watermarks, as one primary use case for open-source models is for users to share and deploy edited versions, as evidenced by almost $200$ thousand models hosted on Hugging Face, with over $200$M downloads.
Hence, having a well-defined and systematically evaluated notion of durability is crucial to guide OSM watermark research.

\section{Durability of Open-source LLM Watermarks}
\label{sec:durability}

Next, we proceed to define the durability requirement more concretely. In~\cref{sec:methods} we will introduce current OSM watermarking methods, and use our durability evaluation setup to systematically evaluate them in \cref{sec:eval}.
To concretize the durability requirement, we have surveyed both the literature and trending Hugging Face models.  
We identified four main categories of modifications: quantization, pruning, merging, and finetuning, among which we select the most prominent methods and parameter settings.

\paragraph{Quantization} 
Model quantization techniques have emerged as a key method to enable the deployment of increasingly large LLMs on memory-constraint commodity hardware. The fundamental idea underlying quantization is to represent (\textit{quantize}) model weights (and activations) in lower-precision data types. We can split popular methods into two categories: \emph{zero-shot} and \emph{optimization-based}.
Zero-shot methods fix the quantization mapping (\textit{buckets}) independently of the model on which they are applied. This makes them computationally inexpensive and a popular choice in consumer libraries (e.g., \textsc{LLM.int8()} \citep{llmint8}, and \textsc{NF4} \citep{qlora} in Hugging Face). Meanwhile, optimization-based methods aim to minimize a reconstruction error assuming a specific model. This includes methods like \textsc{HQQ} \citep{hqq}, which optimizes reconstruction error only over model weights, as well as a range of methods, such as \textsc{GPTQ} \citep{gptq} and \textsc{AWQ} \citep{awq}, that optimize activation reconstructions over an additional calibration dataset.
For evaluating durability, we consider both 8 bits and 4 bits quantized models with different methods.

\paragraph{Pruning} 
While quantization reduces precision across weights, pruning aims to reduce memory requirements by directly removing specific weights completely (\textit{zeroing out}).
Unstructured pruning techniques such as \textsc{Wanda} \citep{wanda}, \textsc{SparseGPT} \citep{sparsegpt}, and \textsc{GBLM} \citep{gblm_pruner} independently remove weights while relying on minimizing a reconstruction error between the pruned weights and the dense weights on a calibration dataset. 
On the contrary, structured pruning methods such as \textsc{Sheared Llama} \citep{sheared_llm} and \textsc{LLM-Pruner} \citep{llm_pruner} aim to remove entire sets of weights (e.g., rows, columns, or layers) jointly (likewise minimizing a reconstruction error).
The advantage of structured pruning is that the resulting model inhibits dense substructures in its weights, allowing for hardware-optimized inference algorithms. 
At the same time, they are usually not zero-shot and require additional finetuning to restore model performance after pruning.  
Given such additional modifications, we will only focus on unstructured pruning methods.

\paragraph{Model merging}
Model merging techniques aim to construct a new model out of a set of base models by combining their individual weights. 
Importantly, previous works \citep{fischer_merging, dataless_merging} have shown that model merging allows for cheaply combining multiple expert models into a single model that maintains task-specific performance. 
Most merging techniques \citep{fischer_merging, dataless_merging, ties_merging, dare_merging} thereby rely on the concept of task vectors and task arithmetic: expert knowledge in LLMs lies in orthogonal directions in weight space and can be directly combined to obtain a vector that joins their respective strengths.

We focus on merging via Spherical Linear Interpolation (\textsc{SLERP}) between the watermarked and original model \citep{mergekit}.
Given the original model $\theta_0$, the watermarked model $\theta_{\text{wm}}$, the angle $\Omega$ between $\theta_0$ and $\theta_{\text{wm}}$ (we set $\Omega := \frac{\pi}{2}$ if $\theta_0$ or $\theta_{\text{wm}}$ is null), and the interpolation parameter $t \in [0,1]$, we consider  
\begin{equation} 
 \text{SLERP}(\theta_{\text{wm}},\theta_0, t) = \frac{\sin[(1-t) \Omega]}{\sin\Omega} \theta_{\text{wm}} + \frac{\sin[t\Omega]}{\sin \Omega} \theta_0.
\end{equation}
Evaluating durability on the \textsc{SLERP} merge with the original model provides both a reproducible setting for comparing different OSM watermarks and a more adversarial scenario than practical applications.  
Indeed, merges are performed on models from the same family, and hence, in the case of a watermarked model, all merged models are normally derived from the watermarked model.  

\paragraph{Finetuning}
Model finetuning is widely used to improve pretrained models on a specific domain, usually via additional training on a domain-specific dataset. 
Besides full model finetuning, which updates all model weights via gradient descent, Low-Rank Adaptation (\textsc{LoRA}) \citep{lora} has emerged as an incredibly popular finetuning method that performs parameter-efficient low-dimensional weight updates. 
Besides introducing domain-specific knowledge, one of the common finetuning use cases is instruction finetuning \citep{instruction_tuning}, where a base model is trained to follow the instruction format of a given Q\&A dataset, enabling its usage as a chat model.

Apart from such supervised finetuning (SFT) methods, Reinforcement Learning (RL)-based finetuning \citep{rlhf, rl_survey} is commonly applied to align models with human preferences or enable more complex reasoning behavior. 
Yet, due to the additional complexity of RL-based finetuning, it is, for open-source models, so far significantly less common than SFT.
Hence, to evaluate watermark durability against finetuning, we focus only on SFT as well as instruction finetuning, both on the full weights of the model and with \textsc{LoRA}.

\section{Current State of Open-source LLM Watermarking} \label{sec:methods}

Next, we introduce the current state of watermarking schemes for open-source models. 
We identify two main categories: schemes that embed the watermark into the model using gradient descent \citep{learnability,rlwatermark}, and those that directly edit the weights \citep{gaussmark, unremovable}.
The latter are less computationally expensive but either require architectural changes to the model~\citep{unremovable} or more compute-intensive detection~\citep{gaussmark}.
For gradient-based methods, there remain unexplored questions about how such methods generalize to different tasks \citep{wapiti} or the viability of statistical guarantees \citep{rlwatermark}. 
As many OSM watermarks are based on generation-time watermarks, we provide a separate introduction in \cref{app:sampling_based_wm}.
We evaluate durability against common modifications of the methods presented here in \cref{sec:eval}.

\paragraph{Distillation-based watermark}
In \citet{learnability}, the authors show that generation-time watermarks \citep{kgw,aar, stanford} can be imprinted into the model weights by distilling the watermark from a teacher model $\theta_0$.
Then, the same watermark detector can be used to detect the watermark in the student model $\theta$.
In the first variant, the teacher model is used in a black-box way to generate watermarked data $\mathcal{D}_{\text{wm}}$, which the student finetunes on using the cross-entropy loss:
\begin{equation}
 \mathcal{L}_{\text{sampling}} (\theta) = \mathbb{E}_{x \sim \mathcal{D}_{\text{wm}}} \left[ \sum_{t=1}^{|x|} -\log p_\theta (x_t | x_{<t}) \right].
\end{equation}
In the white-box variant, the student model is finetuned to mimic the teacher model's next-token distribution, using a loss based on $\text{KL}$-divergence:
\begin{equation}
    \label{eq:distillation_learnability}
 \mathcal{L}_{\text{logit}} (\theta) = \mathbb{E}_{x \sim \mathcal{D}} \left[ \sum_{t=1}^{|x|} \text{KL}(f_w(p_{\theta_0}(.|x_{<t}), \xi_w), p_{\theta}(.|x_{<t})) \right].
\end{equation} 
For evaluating durability, we distill two different generation-time watermarks, \textsc{KGW} and \textsc{KTH} (see \cref{app:sampling_based_wm}).  
We label the corresponding distilled OSM watermarks \KGW and \KTH, respectively.

\paragraph{RL-based watermark}
In \citet{rlwatermark}, the authors propose integrating the watermark into the RLHF pipeline \citep{rlhf} by jointly training the watermark and the watermark detector using reinforcement learning.  
More precisely, given a dataset $\mathcal{D} = \{(x_i,y_i)\}$ of prompts and non-watermarked completions, a watermark detector $D$ parameterized by $\theta^d$, and two models $\theta^0, \theta \in \Theta$, we optimize the following objective using PPO \citep{ppo}:  
\begin{equation}
    \min_{\theta^d, \theta} \mathbb{E}_{(x,y) \sim \mathcal{D}}[D(x,y,\theta^d) - D(x,y(x,\theta),\theta^d)] + \lambda \text{Reg}(\theta, \theta^0).
\end{equation}
 
\paragraph{Weight-editing watermarks}
Both \citet{unremovable} (\unremovable) and \citet{gaussmark} (\gaussmark) propose directly editing the weights of the model without needing gradient descent.

\unremovable introduces a Gaussian noise $\varepsilon = \mathcal{N}(0, \sigma I_{|\Sigma|})$ bias layer in the last projection matrix.
The detector, given a text $x \in \Sigma^*$, computes the following Z-score and performs a one-sided Z-test:
\begin{equation}
    \label{eq:unremovable_detector}
    Z(x, \varepsilon) = \frac{\sum_{t=1}^{|x|} \varepsilon[x_t]}{\sigma|x|}.
\end{equation}
As no prominent open-source model architecture has a bias layer in the last projection matrix, \unremovable requires a modification of the model architecture allowing for simple removal by disabling the respective layer.
Hence, for current architectures, this watermark is not durable---we still include it in our evaluation for completeness.

\gaussmark generalizes \unremovable to target any subset $\theta_r \subset \theta$ of existing model weights (with dimension $d_r \in \mathbb{N}$).
For this we compute $\varepsilon = \mathcal{N}(0, \sigma I_{d_r})$ and consider the model with $\theta_r + \varepsilon$ and all other weights $\theta \backslash \theta_r$ untouched. For detection, we use the following statistic:
\begin{equation}
    \label{eq:gaussmark-detector}
    Z(x, \varepsilon) = \frac{ \varepsilon \cdot \nabla_{\theta_r} \log(p_{\theta}(x)) }{\sigma \lVert \nabla_{\theta_r} \log(p_{\theta}(x)) \rVert _2},
\end{equation}
and also perform a one-sided Z-test. Unlike \unremovable, \gaussmark can be applied to any subset of weights from the model and, crucially, does not necessarily require editing the architecture of the model.
However, it requires a forward and (partial) backward pass for watermark detection.

\section{Evaluation}
\label{sec:eval}

\begin{table}[t]\centering
  \caption{Durability evaluation: TPR at $5\%$ FPR, and median PPL of different watermarked versions of \llama under model modifications. (L) denotes LoRA. The missing values indicate cases where a modification was not applicable due to the watermark's architectural changes.}
  \label{tab:main_results_table}
  \vspace{0.02in}

  \newcommand{\onecol}[1]{\multicolumn{1}{c}{#1}}
  \newcommand{\twocol}[1]{\multicolumn{2}{c}{#1}}
  \newcommand{\threecol}[1]{\multicolumn{3}{c}{#1}} 
  \newcommand{\fourcol}[1]{\multicolumn{4}{c}{#1}}
  \newcommand{\fivecol}[1]{\multicolumn{5}{c}{#1}}
  \newcommand{\sixcol}[1]{\multicolumn{6}{c}{#1}}
  \newcommand{\sevencol}[1]{\multicolumn{7}{c}{#1}}
  \newcommand{\eightcol}[1]{\multicolumn{8}{c}{#1}}
  \newcommand{\ninecol}[1]{\multicolumn{9}{c}{#1}}

  \renewcommand{\arraystretch}{1.2}
  \newcommand{\skiplen}{0.000001\linewidth} 
  \newcommand{\rlen}{0.01\linewidth} 
  \resizebox{\linewidth}{!}{%
  \begingroup 
  \setlength{\tabcolsep}{5pt} %
  \begin{tabular}{lll
    x{1}{2}
    y{2}{1}
    x{1}{2}
    y{2}{1}
    x{1}{2}
    y{2}{1} 
    x{1}{2}
    y{2}{1}
    x{1}{2}
    y{2}{1}
  }

      \toprule
      &&& \twocol{\textsc{KGW-D}} & \twocol{\textsc{KTH-D}} & \twocol{\textsc{\shortstack{Unre-\\movable}}} & \twocol{\textsc{\shortstack{Gauss\\Mark}}} & \twocol{\textsc{\shortstack{KGW-D\\+CTV}}} \\
    \cmidrule{4-5}
    \cmidrule{6-7}
    \cmidrule{8-9}
    \cmidrule{10-11}
    \cmidrule{12-13}

     \threecol{\textbf{Model Modification}}& \textbf{\shortstack{TPR\\@5}}& \textbf{PPL}  & \textbf{\shortstack{TPR\\@5}} & \textbf{PPL}  & \textbf{\shortstack{TPR\\@5}} & \textbf{PPL}  & \textbf{\shortstack{TPR\\@5}} & \textbf{PPL}  & \textbf{\shortstack{TPR\\@5}} & \textbf{PPL} \\
     \midrule
     Unaltered &&&\cellcolor{green!25} 0.99 & 6.63 &\cellcolor{yellow!25} 0.81 & 9.65 &\cellcolor{green!25} 0.98 & 7.47 &\cellcolor{green!25} 0.99 & 7.46 &\cellcolor{green!25} 0.99 & 9.06 \\
    \midrule
    \multirow{5}{*}{Quantization} &  \multirow{2}{*}{8 bits} & \textsc{GPTQ} &\cellcolor{green!25} 0.99 & 6.60 &\cellcolor{yellow!25} 0.85 & 7.70 &\cellcolor{green!25} 0.99 & 7.70 &\cellcolor{green!25} 0.99 & 7.20 &\cellcolor{green!25} 0.96 & 9.20 \\
    && INT8 &\cellcolor{green!25} 0.99 & 6.50 &\cellcolor{red!25} 0.76 & 7.10 &\cellcolor{green!25} 0.99 & 8.00 &\cellcolor{green!25} 0.96 & 7.90 &\cellcolor{green!25} 1.00 & 8.80 \\
    \cmidrule{2-13}
     & \multirow{3}{*}{4 bits} & \textsc{HQQ} &\cellcolor{green!25} 0.99 & 6.80 &\cellcolor{red!25} 0.78 & 7.00 &\cellcolor{green!25} 0.98 & 8.30 &\cellcolor{green!25} 0.95 & 8.40 &\cellcolor{green!25} 0.99 & 9.90 \\
    && \textsc{GPTQ} &\cellcolor{green!25} 0.99 & 7.00 &\cellcolor{red!25} 0.78 & 8.00 &\cellcolor{green!25} 0.99 & 8.20 &\cellcolor{green!25} 0.96 & 8.40 &\cellcolor{green!25} 1.00 & 10.50 \\
    && \textsc{NF4} &\cellcolor{green!25} 0.98 & 6.60 &\cellcolor{red!25} 0.77 & 7.40 &\cellcolor{green!25} 0.99 & 8.60 &\cellcolor{green!25} 0.97 & 8.00 &\cellcolor{green!25} 0.99 & 9.70 \\
    \midrule
    \multirow{6}{*}{Pruning} & \multirow{2}{*}{\textsc{Wanda}} & $\rho = 0.2$ &\cellcolor{green!25} 0.99 & 10.00 &\cellcolor{yellow!25} 0.88 & 11.10 &\cellcolor{red!25} N/A & N/A &\cellcolor{green!25} 0.98 & 8.00 &\cellcolor{green!25} 1.00 & 9.50 \\
    &&$\rho = 0.5$ &\cellcolor{green!25} 1.00 & 8.40 &\cellcolor{red!25} 0.79 & 7.80 &\cellcolor{red!25} N/A & N/A &\cellcolor{green!25} 0.97 & 10.50 &\cellcolor{green!25} 0.99 & 9.50 \\
    \cmidrule{2-13}
    & \multirow{2}{*}{\textsc{GBLM}} & $\rho = 0.2$ &\cellcolor{green!25} 0.99 & 9.30 &\cellcolor{yellow!25} 0.85 & 9.90 &\cellcolor{red!25} N/A & N/A &\cellcolor{green!25} 0.98 & 7.60 &\cellcolor{green!25} 1.00 & 9.10 \\
    && $\rho = 0.5$ &\cellcolor{green!25} 0.98 & 8.00 &\cellcolor{red!25} 0.76 & 8.30 &\cellcolor{red!25} N/A & N/A &\cellcolor{green!25} 0.91 & 10.60 &\cellcolor{green!25} 0.99 & 9.00 \\
    \cmidrule{2-13}
    & \multirow{2}{*}{\textsc{SparseGPT}} & $\rho = 0.2$ &\cellcolor{green!25} 1.00 & 10.10 &\cellcolor{yellow!25} 0.86 & 13.20 &\cellcolor{red!25} N/A & N/A &\cellcolor{green!25} 0.98 & 8.60 &\cellcolor{green!25} 1.00 & 10.30 \\
    && $\rho = 0.5$ &\cellcolor{green!25} 1.00 & 8.80 &\cellcolor{yellow!25} 0.89 & 10.00 &\cellcolor{red!25} N/A & N/A &\cellcolor{green!25} 0.94 & 10.80 &\cellcolor{green!25} 0.99 & 12.90 \\
    \midrule
    \multirow{5}{*}{Merging} & \multirow{5}{*}{\textsc{SLERP}} & $t = 0.1$ &\cellcolor{green!25} 0.98 & 8.80 &\cellcolor{red!25} 0.52 & 8.00 &\cellcolor{green!25} 0.98 & 7.60 &\cellcolor{green!25} 0.97 & 11.50 &\cellcolor{green!25} 0.98 & 12.80 \\
    && $t = 0.3$ &\cellcolor{yellow!25} 0.82 & 7.60 &\cellcolor{red!25} 0.18 & 7.30 &\cellcolor{green!25} 0.98 & 7.40 &\cellcolor{yellow!25} 0.82 & 11.10 &\cellcolor{yellow!25} 0.83 & 10.90 \\
    && $t = 0.5$ &\cellcolor{red!25} 0.50 & 7.70 &\cellcolor{red!25} 0.15 & 6.90 &\cellcolor{green!25} 0.93 & 7.50 &\cellcolor{red!25} 0.68 & 12.00 &\cellcolor{red!25} 0.59 & 11.00 \\
    && $t = 0.7$ &\cellcolor{red!25} 0.17 & 7.10 &\cellcolor{red!25} 0.11 & 7.00 &\cellcolor{yellow!25} 0.83 & 6.70 &\cellcolor{red!25} 0.33 & 10.70 &\cellcolor{red!25} 0.34 & 9.70 \\
    && $t = 0.9$ &\cellcolor{red!25} 0.04 & 7.10 &\cellcolor{red!25} 0.09 & 7.40 &\cellcolor{red!25} 0.35 & 6.80 &\cellcolor{red!25} 0.10 & 10.00 &\cellcolor{red!25} 0.09 & 10.10 \\
    \midrule
    \multirow{8}{*}{Finetuning}&\multirow{4}{*}{\textsc{\shortstack[l]{Open\\WebText}}}& 500 (L) &\cellcolor{red!25} 0.76 & 6.00 &\cellcolor{red!25} 0.14 & 5.70 &\cellcolor{green!25} 0.94 & 7.30 &\cellcolor{yellow!25} 0.86 & 6.90 &\cellcolor{red!25} 0.72 & 7.70 \\
    && 2500 (L) &\cellcolor{red!25} 0.52 & 5.50 &\cellcolor{red!25} 0.13 & 5.80 &\cellcolor{red!25} 0.62 & 7.40 &\cellcolor{red!25} 0.67 & 7.20 &\cellcolor{red!25} 0.56 & 7.00 \\
    \cmidrule{3-13}
     &  & 500 &\cellcolor{red!25} 0.35 & 5.70 &\cellcolor{red!25} 0.09 & 5.30 &\cellcolor{red!25} 0.24 & 7.10 &\cellcolor{red!25} 0.48 & 7.70 &\cellcolor{red!25} 0.34 & 7.30 \\
    && 2500 &\cellcolor{red!25} 0.22 & 5.50 &\cellcolor{red!25} 0.10 & 5.70 &\cellcolor{red!25} 0.23 & 7.00 &\cellcolor{red!25} 0.29 & 7.10 &\cellcolor{red!25} 0.22 & 7.40 \\
    \cmidrule{2-13}
    &\multirow{4}{*}{\textsc{\shortstack[l]{Open\\MathInstruct}}}& 500 (L) &\cellcolor{green!25} 0.99 & 6.40 &\cellcolor{red!25} 0.58 & 5.60 &\cellcolor{green!25} 0.98 & 6.90 &\cellcolor{green!25} 0.97 & 7.20 &\cellcolor{green!25} 0.99 & 8.10 \\
    && 2500 (L) &\cellcolor{green!25} 0.99 & 6.50 &\cellcolor{red!25} 0.47 & 5.50 &\cellcolor{green!25} 0.95 & 7.00 &\cellcolor{green!25} 0.96 & 7.20 &\cellcolor{green!25} 0.98 & 7.90 \\
    \cmidrule{3-13}
    &  & 500 &\cellcolor{red!25} 0.75 & 4.90 &\cellcolor{red!25} 0.18 & 5.00 &\cellcolor{red!25} 0.67 & 6.40 &\cellcolor{red!25} 0.69 & 6.40 &\cellcolor{yellow!25} 0.87 & 7.20 \\
    && 2500 &\cellcolor{red!25} 0.55 & 4.60 &\cellcolor{red!25} 0.12 & 4.50 &\cellcolor{red!25} 0.51 & 5.90 &\cellcolor{red!25} 0.66 & 5.80 &\cellcolor{red!25} 0.69 & 5.80 \\

     \bottomrule

     \end{tabular}
    
  \endgroup
  }
\end{table}

In this section, we present the results of our experimental evaluation of the durability of OSM watermarks (\cref{sec:methods}) against common model modifications (\cref{sec:durability}), with our results highlighting that durable OSM watermarking is still an open challenge.
We defer omitted experimental details to \cref{app:experimental_details} and present additional in-depth results for each model modification in \cref{sec:deep_dives}. 

\paragraph{Methods}
We evaluate \KGW with $\delta = 2$, $\gamma = 0.25$, and $k = 1$, and \KTH with key size 256 and no key shift.
We do not evaluate distilled AAR~\citep{aar} as it highly degrades text quality~\citep{learnability}.
For \unremovable, we set $\sigma = 0.6$, and do not evaluate it against pruning, as current pruning methods assume no architectural changes. 
For \gaussmark we set $\sigma = 0.018$ and apply it to the up-projection matrix (\texttt{up\_proj}) of the MLP layer in the $31$st attention block.
We omit the RL-based watermark, as we were unable to train it to a sufficient text quality.

\paragraph{Variant: targeted distillation}
In addition to methods from prior work, we evaluate an additional variant of \KGW that leverages \emph{contrastive task vectors} (\textsc{CTV}) \citep{task_vector, ctv} in an aim to improve durability.
Namely, we first apply \KGW to $\theta_0$ to obtain $\theta_1$. 
Then, we finetune $\theta_1$ on a broad-domain dataset (\textsc{OpenWebText}, see \cref{app:experimental_details}) to obtain $\theta_2$ where the watermark has been removed.
Let $\tau$ denote the following boolean mask of the model weights: 
\begin{equation}
    \label{eq:ctv}
    \tau = |\theta_1 - \theta_0| > |\theta_2 - \theta_0|.
\end{equation}
Intuitively, weights where $\tau$ is false were leveraged to remove the watermark---we aim to avoid relying on such weights in our final model by (again) applying \KGW to $\theta_0$ only where $\tau$ is true. 

\paragraph{Model modifications}
For each modification from \cref{sec:durability}, we evaluate a range of representative settings.
For quantization, we use 4-bit and 8-bit variants, using \textsc{HGG}, \textsc{LLM.int8()}, \textsc{GPTQ}, and \textsc{AWQ} methods---going below $4$ bits significantly degraded text quality in our experiments.
We evaluate pruning with sparsity ratios $\rho$ $\in \{0.2,0.5\}$ using unstructured pruning methods \textsc{Wanda}, \textsc{GBLM}, and \textsc{SparseGPT}. 
For merging, we merge the watermarked model with the base model using \textsc{SLERP} with interpolation ratios $t \in \{0.1,0.3,0.5,0.7,0.9\}$.
Finally, we consider full finetuning and parameter-efficient \textsc{LoRA}, both on broad-domain \textsc{OpenWebText}~\citep{openwebtext} (Reddit, completions) and task-specific \textsc{OpenMathInstruct}~\citep{openmathinstruct} (math, instruction tuning).

We use the \llama model in all experiments, watermark it, apply the model modification, and generate $100$ completions of length $200$ by prompting with the first $50$ tokens of each entry in the RealNewsLike split of C4~\citep{c4}, as in prior work~\citep{kgw}.  
For each completion, we evaluate the watermark strength (TPR at $5$\% FPR, see~\cref{app:experimental_details}) and median quality (PPL using \textsc{LLama3-8B}) to ensure that our modifications sufficiently retain text quality.

\paragraph{Results: OSM watermarks lack durability}
In \cref{tab:main_results_table}, we present our main results. 
Colors correspond to different TPR ranges (\textcolor{green!70!black}{green}: above $0.9$, \textcolor{yellow!70!black}{yellow}: between $0.8$ and $0.9$, \textcolor{red!70!black}{red}: below $0.8$).

First, we observe that nearly all tested schemes are durable to quantization, irrespective of the quantization method, and even at $4$ bits. 
Similar results hold for pruning, where all schemes are highly durable for $\rho=0.2$ and remain significantly durable up to a sparsity ratio of $0.5$. 
This intuitively follows from the fact that both quantization and pruning directly aim to minimize the distortion between quantized and original model, thereby also preserving the embedded watermark.

For merging with the unwatermarked model using \textsc{SLERP}, we can observe that weight-editing watermarks are, on average, more durable than distillation-based ones.
For \unremovable, these good results are expected, as merging the bias layer with the null vector (i.e., the bias layer of the unwatermarked model) is equivalent to applying the same noise $\varepsilon$ but with a scaled standard deviation
\begin{equation}
    \varepsilon_{\text{SLERP}}(t) = \sin[(1-t)\frac{\pi}{2}] \varepsilon.
\end{equation}
Still, for $t \geq 0.7$ all methods struggle to retain the watermark.
Interestingly, we see that \textsc{KGW-D+CTV} is slightly more durable than \KGW, suggesting that the contrastive task vectors can indeed improve durability, presumably as they localize the watermark to fewer parameters---however, this modification is ultimately ineffective, as it does not improve strength across other model modifications.

Most importantly, we find that for full finetuning on \textsc{OpenWebText}, \textbf{none of the tested watermarking schemes are durable}: after only $500$ steps of gradient descent, the TPR drops significantly to a point where the watermark seizes to be useful.  
Similar results hold for LoRA finetuning on \textsc{OpenWebText}, where only \unremovable remains high TPR after $500$ steps before dropping significantly after $2500$ steps of finetuning. As in the other cases, this can be explained by the architectural modifications of \unremovable: LoRA finetuning does not directly modify the (usually not present) last layer bias and, therefore, cannot directly modify the watermarked part of the model.

We extend these results by including finetuning experiments with the domain-specific dataset \textsc{OpenMathInstruct}, modeling a realistic use-case where a user would finetune an OSM on an expert task.
Across all schemes, we find higher durability compared to \textsc{OpenWebText}, especially for \textsc{LoRA}.
Note that we still evaluate the watermark strength on the general domain C4 test set---as we show in \cref{sec:deep_dives}, there is a significant drop in watermark strength when evaluated on the math domain. 
This points to an interesting phenomenon of \emph{domain-specificity} of OSM watermarks: watermark strength more quickly degrades on domains specifically targeted by finetuning. 

Overall, we conclude that while prior work proposed a range of OSM watermarking methods with varying tradeoffs, no method is currently sufficiently durable.
Given the real-world prominence of such model modifications, ensuring durability against them remains an open and critical challenge for future research---here our proposed evaluation setup provides an easy way to compare future work.

\section{Improving OSM Watermark Durability}
\label{sec:wm_from_scratch}
 
In this section, we extend current distillation-based OSM methods introduced in \cref{sec:methods} by significantly increasing the distillation dataset size and further explore a variant that starts from a randomly initialized model (\emph{distillation pretraining}), as opposed to the standard application on top of an already pretrained model~\citep{learnability}.
As a proof of concept, we show that on a \textsc{GPT-2} architecture, distilling on a large training set significantly improves watermark durability.
Moreover, we show that distillation pretraining and standard distillation finetuning exhibit complementary behaviors: distillation pretraining is more durable against specific task finetuning, whereas standard distillation is more durable against broad-domain finetuning.
We further expand on our results in \cref{app:from_scratch_extended}.

\paragraph{Experimental details}
We perform the distillation of the \textsc{KGW} watermark with $\delta=2, \gamma=0.25$, and $k=1$ on a \textsc{GPT-2}-based architecture.
We use the following setup: \textsc{KGW-D (Pretrained)} trains a model with random initialization $\theta \in \Theta$ using \citet{learnability} (see \cref{eq:distillation_learnability}) with $\approx$9B tokens.
For \textsc{KGW-D (Long)}, we finetune an already pretrained model ($\theta_0$) with the same distillation approach, also with $\approx$9B tokens.
As a reference, we also distill the watermark using the same hyperparameters as in \citet{learnability}, \ie with only $\approx 40$ million tokens on top of $\theta_0$ (\textsc{KGW-D (Standard)}).
To evaluate watermark strength, we use the same setup as in~\cref{sec:eval} but with $1000$ completions instead of $100$.
To evaluate durability to model modifications, as in \cref{sec:eval}, we finetune on both broad-domain \textsc{OpenWebText} and task-specific \textsc{OpenMathInstruct} datasets.

\paragraph{Distillation on more tokens is more durable}
In \cref{tab:scratch_table}, we see that both \textsc{KGW-D (Pretrained)} and \textsc{KGW-D (Long)} are significantly more durable against finetuning compared to \textsc{KGW-D (Standard)}.
This confirms our prior intuition that OSM watermark durability scales with the extent of the model's exposure to watermarked text during training.
However, while the results are promising when finetuning on \textsc{OpenMathInstruct}, finetuning on a broad-domain dataset such as \textsc{OpenWebText} still significantly deteriorates the watermark.

\begin{table}[t]\centering
    \caption{Durability evaluation: TPR at $1\%$ and $5\%$ FPR, and median PPL of different watermarked versions of \textsc{GPT-2} under finetuning on either \textsc{OpenWebText} or \textsc{OpenMathInstruct}.}
    \label{tab:scratch_table}
    \vspace{0.02in}
  
    \newcommand{\onecol}[1]{\multicolumn{1}{c}{#1}}
    \newcommand{\twocol}[1]{\multicolumn{2}{c}{#1}}
    \newcommand{\threecol}[1]{\multicolumn{3}{c}{#1}} 
    \newcommand{\fourcol}[1]{\multicolumn{4}{c}{#1}}
    \newcommand{\fivecol}[1]{\multicolumn{5}{c}{#1}}
    \newcommand{\sixcol}[1]{\multicolumn{6}{c}{#1}}
    \newcommand{\sevencol}[1]{\multicolumn{7}{c}{#1}}
    \newcommand{\eightcol}[1]{\multicolumn{8}{c}{#1}}
    \newcommand{\ninecol}[1]{\multicolumn{9}{c}{#1}}
  
    \renewcommand{\arraystretch}{1.2}
    \newcommand{\skiplen}{0.000001\linewidth} 
    \newcommand{\rlen}{0.01\linewidth} 
    \resizebox{\linewidth}{!}{%
    \begingroup 
    \setlength{\tabcolsep}{5pt} %
    \begin{tabular}{lll
      x{1}{2}
      x{1}{2}
      y{2}{1}
      x{1}{2}
      x{1}{2}
      y{2}{1}
      x{1}{2}
      x{1}{2}
      y{2}{1} 
    }

        \toprule
        &&& \threecol{\shortstack{\textsc{KGW-D}\\\textsc{(Pretrained)}}} & \threecol{\shortstack{\textsc{KGW-D}\\\textsc{(Long)}}} & \threecol{\shortstack{\textsc{KGW-D} \\ (\textsc{Standard})}} \\
      \cmidrule{4-12}

      \threecol{\textbf{Model Modification}}& \textbf{\shortstack{TPR\\@1}} & \textbf{\shortstack{TPR\\@5}}& \textbf{PPL}  & \textbf{\shortstack{TPR\\@1}} & \textbf{\shortstack{TPR\\@5}}& \textbf{PPL} & \textbf{\shortstack{TPR\\@1}} & \textbf{\shortstack{TPR\\@5}}& \textbf{PPL} \\
       \midrule
       Unaltered &&& \cellcolor{green!25}1.00 & \cellcolor{green!25}1.00 & 34.4& \cellcolor{green!25}1.0 & \cellcolor{green!25}1.0 & 31.8 & \cellcolor{green!25}0.99 & \cellcolor{green!25}1.00 & 30.7 \\
      \midrule
      \multirow{4}{*}{Finetuning} &\multirow{2}{*}{\textsc{OpenWebText}}& 500 & \cellcolor{yellow!25}0.89 & \cellcolor{green!25}0.97 & 25.3 & \cellcolor{green!25}0.90 & \cellcolor{green!25}0.97 & 23.6 & \cellcolor{red!25}0.74 & \cellcolor{yellow!25}0.87 & 24.1 \\
      && 2500 & \cellcolor{red!25}0.57 & \cellcolor{red!25}0.79 & 25.1 & \cellcolor{red!25}0.63 &\cellcolor{yellow!25} 0.84 & 23.5 & \cellcolor{red!25}0.47 & \cellcolor{red!25}0.73 & 23.6 \\
      \cmidrule{2-12}
      &\multirow{2}{*}{\textsc{OpenMathInstruct}}& 500 & \cellcolor{green!25}0.98 & \cellcolor{green!25}0.99 & 24.6 &\cellcolor{green!25} 0.95 & \cellcolor{green!25}0.98 & 21.1 & \cellcolor{yellow!25}0.80 & \cellcolor{yellow!25}0.89 & 21.4 \\
      && 2500 & \cellcolor{green!25}0.91 & \cellcolor{green!25}0.96 & 22.6 & \cellcolor{yellow!25}0.81 & \cellcolor{green!25}0.91 & 19.4 & \cellcolor{red!25}0.48 & \cellcolor{red!25}0.66 & 19.1 \\
      \bottomrule
  
       \end{tabular}
      
    \endgroup
    }
  \end{table}

\begin{figure}[t]
    \centering
    \includegraphics[width=0.95\textwidth]{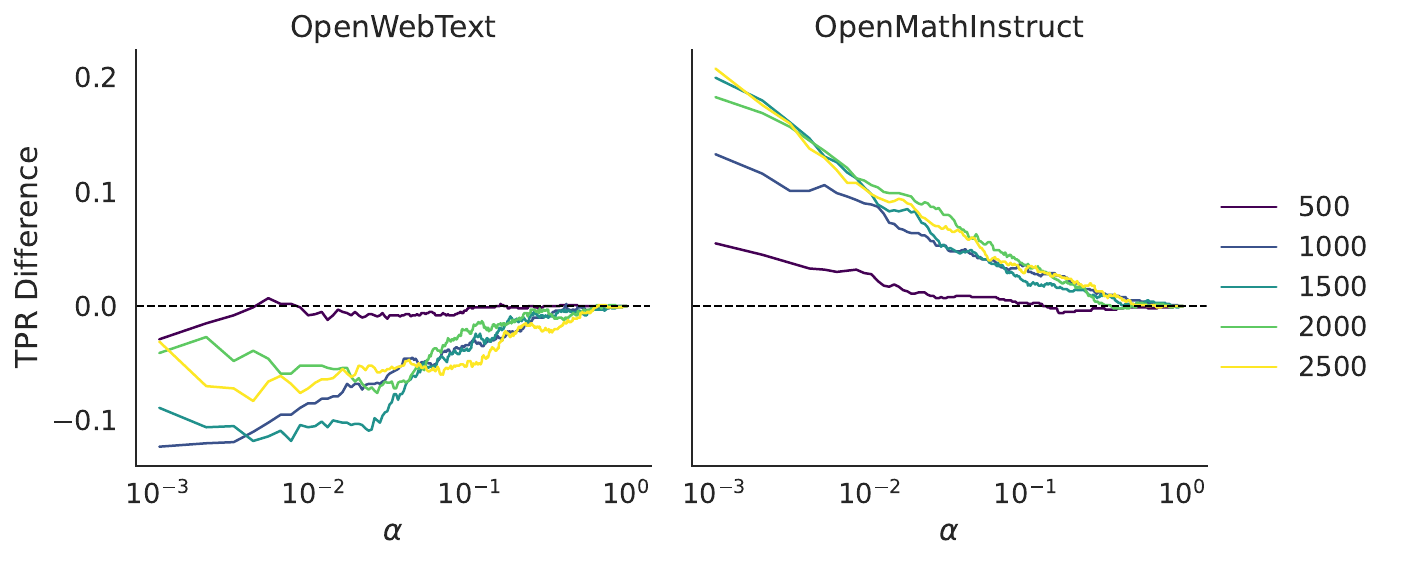}
    \caption{Evaluation of the TPR difference between \textsc{KGW-D (Pretrained)} and \textsc{KGW-D (Long)} when finetuned (as a model modification) on either \textsc{OpenWebText} or \textsc{OpenMathInstruct}.}
    \label{fig:ft_tpr_gpt2_scratch}
\end{figure}

\paragraph{Distillation at pretraining is task-aware}
In \cref{fig:ft_tpr_gpt2_scratch} we present more granular results,
aiming to decouple the effects of training a randomly initialized model and simply increasing the number of training tokens.
Namely, we show the difference of TPR between \textsc{KGW-D (Pretrained)} and \textsc{KGW-D (Long)} across different rejection rates and (model modification) finetuning steps.
We observe that the pretraining distillation approach is slightly worse at preserving the watermark when later finetuned on a broad-domain dataset (at most 10\% TPR difference) but better at preserving the watermark when finetuned on a task-specific dataset (up to 20\% TPR increase).

Perhaps more interestingly, as we evaluate the TPR on C4 prompt completions, \ie a broad-domain/general task, we conclude that the distillation pretraining watermarked model may exhibit a form of \emph{task-awareness}: If the model is finetuned on unwatermarked data from a specific task, it will not un-learn the watermark on other tasks.
We hypothesize that this robustness is due to the model only ever being trained on watermarked text, i.e., never seeing unwatermarked text before the model modification stage.
For the standard distillation watermark, even when the number of training tokens is increased to match the pretraining case, we do not observe this behavior, even though it is, interestingly, slightly more durable against finetuning on a general dataset.

\paragraph{Limitations}
Our results suggest both that the way the model distills the watermark matters, as evidenced by the difference between distillation during pretraining and simple finetuning, and that increasing the training set size can indeed improve durability.  
However, the practicality of these effects is limited, as (1) even with much larger dataset sizes, the increase in durability is often insufficient for the watermark to remain effective against prolonged downstream finetuning, and (2) scaling this approach to real-world models is expensive.  
This also makes it unclear how much our results on the comparatively small \textsc{GPT-2} architecture generalize to larger and more capable models.  

\section{Conclusion}
In this work, we studied the problem of watermarking open-source LLMs.  
Noting the fragmentation in prior work on this topic, we laid the foundation for systematic future study of OSM watermarking.  
We revisited the requirements for generation-time LLM watermarking, discussed how they apply in the open-source setting, introduced a novel requirement, \emph{durability}, and applied a new systematic evaluation procedure to existing OSM watermarks.  
Finding that none of the current watermarks are durable, we proposed scaling watermark distillation methods, highlighting their benefits and limitations.  
We hope our work can kick-start progress on this crucial yet challenging problem.

\bibliography{references}
\bibliographystyle{iclr2025_conference}
\vfill
\clearpage

\message{^^JLASTREFERENCESPAGE \thepage^^J}

\ifincludeappendixx
	\newpage
	\appendix
	\onecolumn 
	\section{Description of Generation-time Watermarks}
\label{app:sampling_based_wm}

In this section, we introduce a high-level description of two generation-time watermarks, \textsc{KGW} \citep{kgw} and \textsc{KTH} \citep{stanford}. 

\paragraph{KGW watermark}
\textsc{KGW} watermark \citep{kgw} works by partitioning, at each step of the generation, the vocabulary into a Red and Green subset using the private key $\xi$ and summing the hashes of the $k$ previous tokens.  
The Green subset has a size of $\gamma |\Sigma|$, with $\gamma \in [0,1]$.  
Logits of the tokens in the Green subset are boosted by $\delta > 0$, making them more likely to be sampled.  
The watermark detector works by performing a binomial test on the number of Green tokens in the text.  

\paragraph{KTH watermark}
\textsc{KTH} watermark \citep{stanford} is parametrized by a key length $n_{key} \in \mathbb{N}$ and a key $\xi \in [0,1]^{|\Sigma| \times n_{key}}$, where each entry $\xi_k \in [0,1]^{|\Sigma|}$ is uniformly distributed in $[0,1]$.
At each step $t$ of the generation, given a probability distribution $p_t$ over $\Sigma$, the next token is chosen as the $\arg\max$ of $(\xi_{(t \mod n_{key})})^{p_t}$.
Additionally, to allow multiple generations given a fixed prompt, the key is randomly shifted by a constant before generating a new sequence.
Finally, given a text $x$, detection works by performing a permutation test using the minimum Levenshtein distance of the alignment cost $d(x, \xi) = \sum_{t=1}^{|x|} \log(1 - \xi_{(t \mod n_{key}, x_t)})$.

\section{Experimental Details}
\label{app:experimental_details}

In this section, we provide an in-depth list of the parameters used for the model modifications (\cref{sec:durability}) and the watermarking schemes (\cref{sec:methods}) that we use in our evaluation in~\cref{sec:eval}.
For all watermarking schemes, we use the same \llama as our base unwatermarked model. 

\paragraph{Watermarking schemes}
For distillation-based generation-time watermarks, we use the distillation loss from \cref{eq:distillation_learnability} with \textsc{OpenWebText} as $\mathcal{D}$.  
We distilled the watermark using the same hyperparameters as in \citet{learnability}: a batch size of $64$, with $512$ tokens per input, a learning rate of 1e-5, the AdamW optimizer \citep{adamw} with $(\beta_1, \beta_2) = (0.9,0.999)$, and no weight decay.  
For \KGW, we use $\delta=2$, $\gamma=0.25$, and $k=1$.  
For \KTH, we use $n_{key} = 256$ with no key shift. 

For \textsc{KGW-D+CTV}, we first distill \textsc{KGW} with the parameters described above on the full model.
Then, we finetune the watermarked model on \textsc{OpenWebText} for 2500 steps with cross-entropy loss, batch size of $64$, $512$ tokens per input, a learning rate of $1\text{e-}5$, the Adafactor optimizer with a cosine learning rate decay, and a linear warmup for the first $500$ steps. 
We then compute the contrastive task vector (\cref{eq:ctv}) and learn \textsc{KGW} on the selected weights.

For the \unremovable watermark, we set the standard deviation to $\sigma=0.6$ to ensure sufficient power in the unaltered watermarked model while not degrading the model's quality too much.  
For \gaussmark, as suggested in \citet{gaussmark}, we perform a grid search to find the best layer and standard deviation to balance watermark power and quality degradation.  
We find the optimal layer to be the 31st MLP up\_proj layer with a standard deviation of $\sigma=0.018$.

\paragraph{Model modifications}
For all quantization and pruning methods, we use the default hyperparameters suggested for each technique. 
For finetuning on \textsc{OpenWebText}, we use a batch size of $32$, with $512$ tokens per input, a learning rate of $1\text{e-}5$, the Adafactor optimizer with a cosine learning rate decay, and a linear warmup for the first $500$ steps. 
For finetuning on \textsc{OpenMathInstruct}, we introduce two new instruction tokens and use the same settings but with a maximum of $2048$ tokens per input to accommodate the entire math problem and solution. 
For both finetuning tasks, we use the same \textsc{LoRA} adapter with a low-rank dimension of $16$ and an alpha of $32$. 
Moreover, we apply the \textsc{LoRA} adapter only to the following layers: v\_proj, k\_proj, o\_proj, and q\_proj.

\paragraph{Metrics}
As our main metric in~\cref{sec:eval} we use the true positive rate of the watermark detector, evaluated at a false positive rate of $5\%$.
While we believe this FPR level is high for fully practical applications, it is both a common evaluation setting in prior watermarking literature, and more importantly, calibrated to the current strength of OSM watermarks.
Ideally, OSM watermarks would advance to a level where this metric is not useful anymore as the corresponding TPRs would be close to $1$ for many methods, and evaluations would focus on lower, more practical FPR levels.

\section{Additional Results on Watermark Durability}
\label{sec:deep_dives}

In this section, we analyze the ROC curves (Experimental TPR versus FPR) to evaluate in detail the durability of the watermarks against all common model modifications (\cref{ssec:roc_curves}).  
We confirm the observation from \cref{sec:eval} that current OSM watermarks lack durability. 
In \cref{ssec:task_specific_eval}, we measure the durability of the watermark on a specific task when finetuning is performed on the same task.  

\subsection{ROC Curves of OSM Watermarks Against Common Model Modifications}
\label{ssec:roc_curves}

Here, we extend the results from \cref{sec:eval} by presenting ROC curves for all schemes tested under all common model modifications identified in \cref{sec:durability}. 

\begin{figure}[t]
    \centering
    \includegraphics[width=\textwidth]{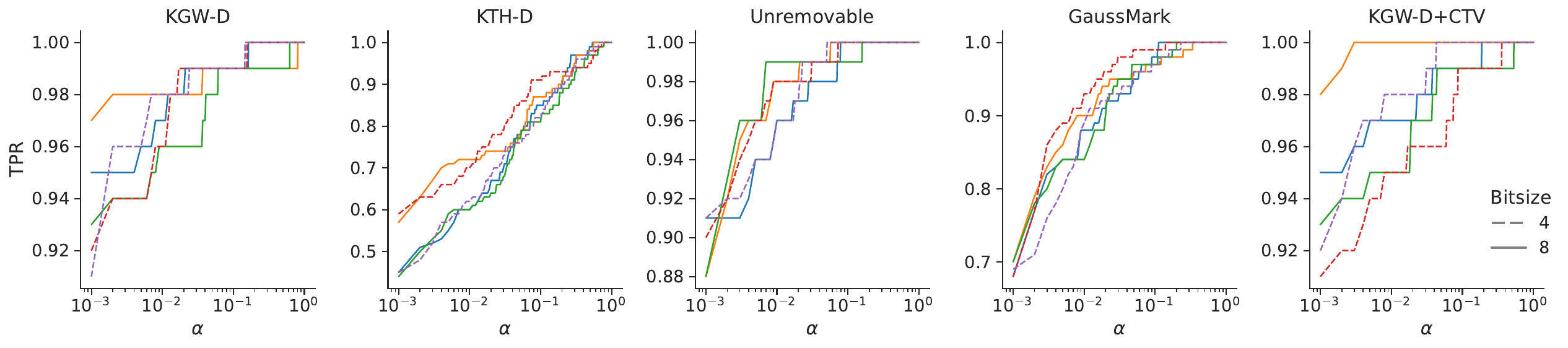}
    \caption{Evolution of different watermark TPRs against multiple quantization methods. Each color corresponds to a different quantization method. The rejection rate $\alpha$ is in logarithmic scale for clarity.}
    \label{fig:quantization_plot}
    \vspace{-0.2em}
\end{figure}

\paragraph{Quantization}
In \cref{fig:quantization_plot}, we see no visible difference between 4 bits and 8 bits quantization in the empirical TPR of different watermarking schemes.
This suggests that quantization is not a challenge for current OSM watermarks.
Intuitively, this is an expected result as most quantization techniques try to preserve the model performance as much as possible when quantizing the model.
Hence, by minimizing the impact on model performance, quantization also minimizes the impact on the watermark.

\begin{figure}[t]
    \centering
    \includegraphics[width=\textwidth]{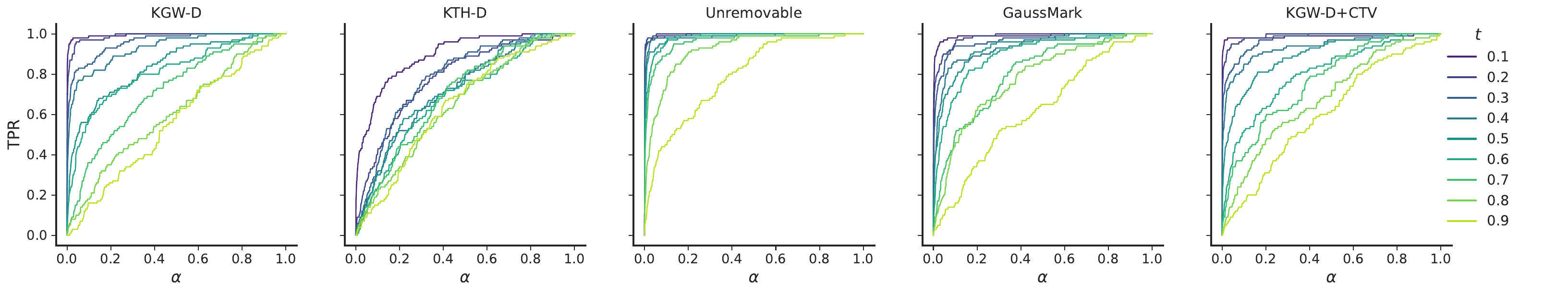}
    \caption{Evolution of different watermark TPRs for different \textsc{SLERP} interpolation levels $t$.}
    \label{fig:merging_tpr}
    \vspace{-0.2em}
\end{figure}

\paragraph{Merging}
Similarly, in \cref{fig:merging_tpr}, we see the ROC curve for different values of the \textsc{SLERP} interpolation parameter and different watermarking schemes.
We see that weight-editing watermarks are more durable against merging.
As explained in \cref{sec:eval}, this is an expected result.

\begin{figure}[t]
    \centering
    \includegraphics[width=\textwidth]{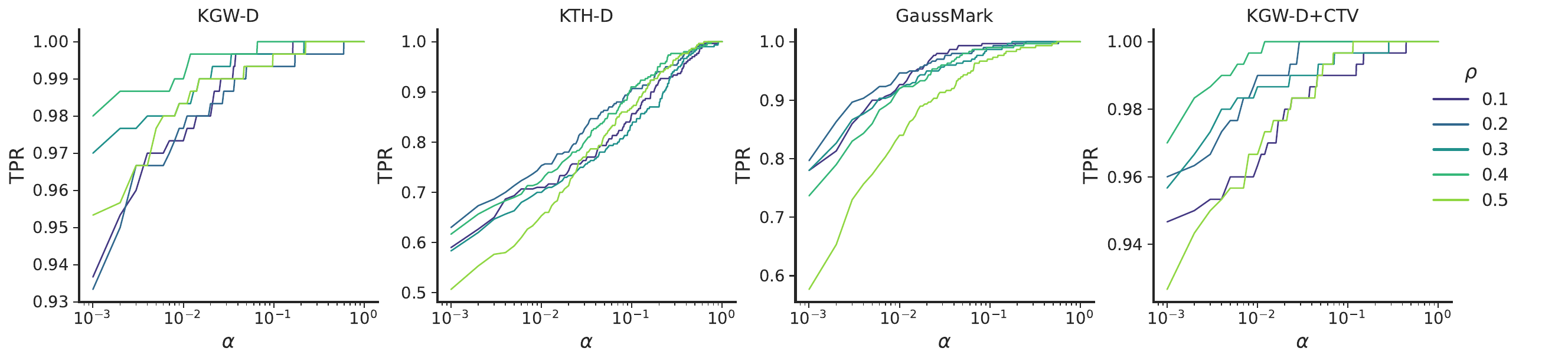}
    \caption{Evolution of different watermark TPRs averaged over three pruning techniques (\textsc{Wanda}, \textsc{GBLM}, and \textsc{SparseGPT}) at different sparsity ratios $\rho$. The rejection rate $\alpha$ is in logarithmic scale.}
    \label{fig:pruning_tpr}
    \vspace{-0.2em}
\end{figure}

\paragraph{Pruning}
In \cref{fig:pruning_tpr}, we see the ROC curve for different values of sparsity $\rho$ and different watermarking schemes.  
For $\rho > 0.5$, the model quality is too low to be usable; hence, we do not compute the TPR for higher sparsity ratios.  
We see that for most schemes tested, the watermark is durable against pruning, even for high sparsity ratios.  
As with quantization, this is an expected result.  
With unstructured pruning, the objective is to find the sparse weights that minimize the distortion in the dense model activations.  
By minimizing such distortion, pruning techniques also preserve the watermark.

\begin{figure}[t]
    \centering
    \includegraphics[width=\textwidth]{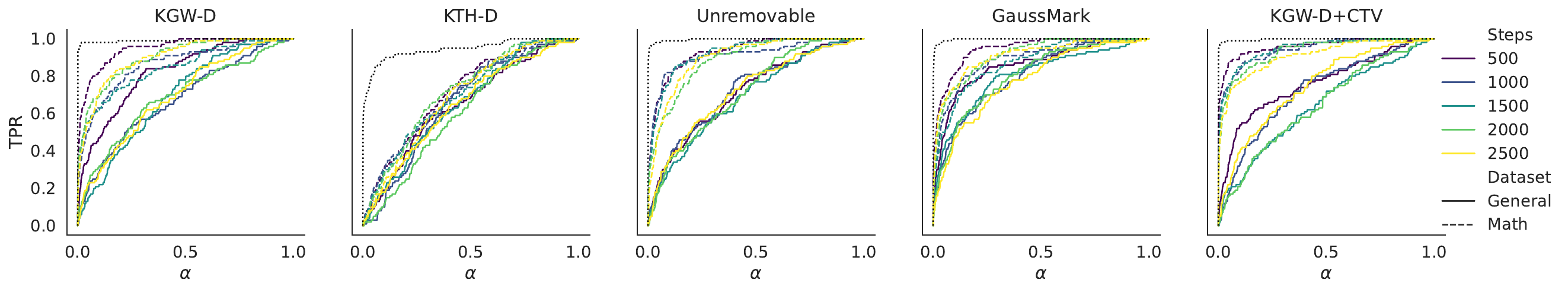}
    \caption{TPR for differen finetuning lengths on either \textsc{OpenWebText} (General) and \textsc{OpenMathInstruct} (Math). The black dotted line corresponds to the unaltered watermark model TPR.}
    \label{fig:ft_tpr}
    \vspace{-0.2em}
\end{figure}

\paragraph{Finetuning}
In \cref{fig:ft_tpr}, we see the ROC curve for full finetuning on either \textsc{OpenWebText} or \textsc{OpenMathInstruct}.
The conclusion is similar as the one from \cref{tab:main_results_table}: the watermark is not durable against finetuning even for a few steps.
This is unsurprising as with finetuning, we want the model to learn the distribution of the training dataset.
Hence, because the training dataset is not watermarked, its distribution significantly differs from the model's previously learned distribution.
Therefore, finetuning effectively bridges the gap between these two distributions, removing the watermark.
This is why, for most schemes, finetuning on a specific domain (Math) does not necessarily remove the watermark as much as finetuning on a broad/general domain.

\subsection{Watermark Durability on Expert Tasks}
\label{ssec:task_specific_eval}

Here, we evaluate the ability of a watermarked open-source model to retain the watermark signal on newly learned expert tasks, e.g., math.
We use the same watermarks as in \cref{sec:eval}.

\paragraph{Experimental details}
We first instruction-finetune a watermarked model on \textsc{OpenMathInstruct} to teach the model how to solve math questions.  
We use the same hyperparameters as in \cref{sec:eval}.  
Then, instead of measuring the watermark durability on broad C4 prompt completions, we consider $200$-token answers to math questions.  
This is closer to a practical scenario: if a user finetunes a given model on a specific task, it is expected that the model will be used for that task as well, hence if the watermark is not durable, most text produced by this model in practice will effectively not be watermarked.

\begin{figure}[!t]
    \centering
    \includegraphics[width=\textwidth]{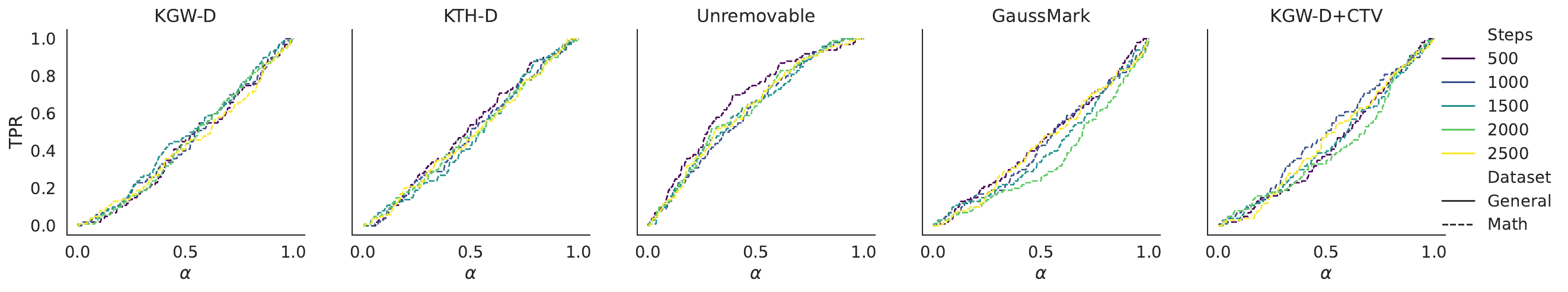}
    \caption{TPR for different lengths of finetuning on \textsc{OpenMathInstruct} (Math). The watermark is evaluated on answers from GSM-8K.}
    \label{fig:ft_tpr_matheval}
    \vspace{-0.2em}
\end{figure}

\paragraph{OSM Watermarks do not transfer to new domains}
In \cref{fig:ft_tpr_matheval}, we see that none of the tested watermarks are durable when finetuned and evaluated on a specific domain.
This contrasts with the evaluation from \cref{tab:main_results_table}, where we show that finetuning on a specific domain but evaluating the watermark strength on a general domain does not lower the watermark strength as significantly.
This highlights a crucial limitation of the durability of current watermarks for open-source models, and suggests that the watermark strength of the task-specific watermarked model should also be evaluated on similar task-specific datasets, which has been overlooked in previous works \citep{wapiti}.

\section{Influence of the Training Length on Distillation-based Watermark Durability}
\label{app:from_scratch_extended}

In this section, we extend the experiment from \cref{sec:wm_from_scratch} and specifically ask whether scaling the number of tokens when distilling the watermark necessarily improves durability against finetuning.

\paragraph{Experimental details}
We again perform the distillation of the \textsc{KGW} watermark with $\delta=2, \gamma=0.25$, and $k=1$ on a \textsc{GPT-2} pretrained model.  
We distill the watermark for a different number of steps, where each step consists of approximately $60$ thousand tokens.  
We then evaluate the watermark strength every $20{,}000$ steps (\ie approximately $1.4$B tokens).  
To evaluate the watermark strength, we use the same setup as in \cref{sec:eval} but with $500$ completions instead of $100$.  
To evaluate durability against finetuning, we finetune on both broad-domain \textsc{OpenWebText} and task-specific \textsc{OpenMathInstruct}.  

\paragraph{Durability does not scale with distillation training length} 
In \cref{fig:ft_tpr_gpt2_ckpts}, we plot both the TPR at 1\% and 5\% FPR for the different models tested.  
We see that the watermarked model that has been distilled for $20$ thousand steps is actually more durable than the one distilled for only $2500$ steps (\textsc{KGW-D (Standard)} from \cref{tab:scratch_table}), but also more durable than the ones distilled for longer (with up to a 10\% TPR@1 difference).  
It also seems that, as the distillation training length increases, its impact on durability plateaus, as all TPR curves from $80$ thousand steps onward are very similar.  
These results confirm the limitations highlighted in \cref{sec:wm_from_scratch}: increasing the training set size, up to a point, improves durability, yet it is still insufficient for the watermark to remain effective against prolonged downstream finetuning.

\begin{figure}[t]
    \centering
    \includegraphics[width=0.95\textwidth]{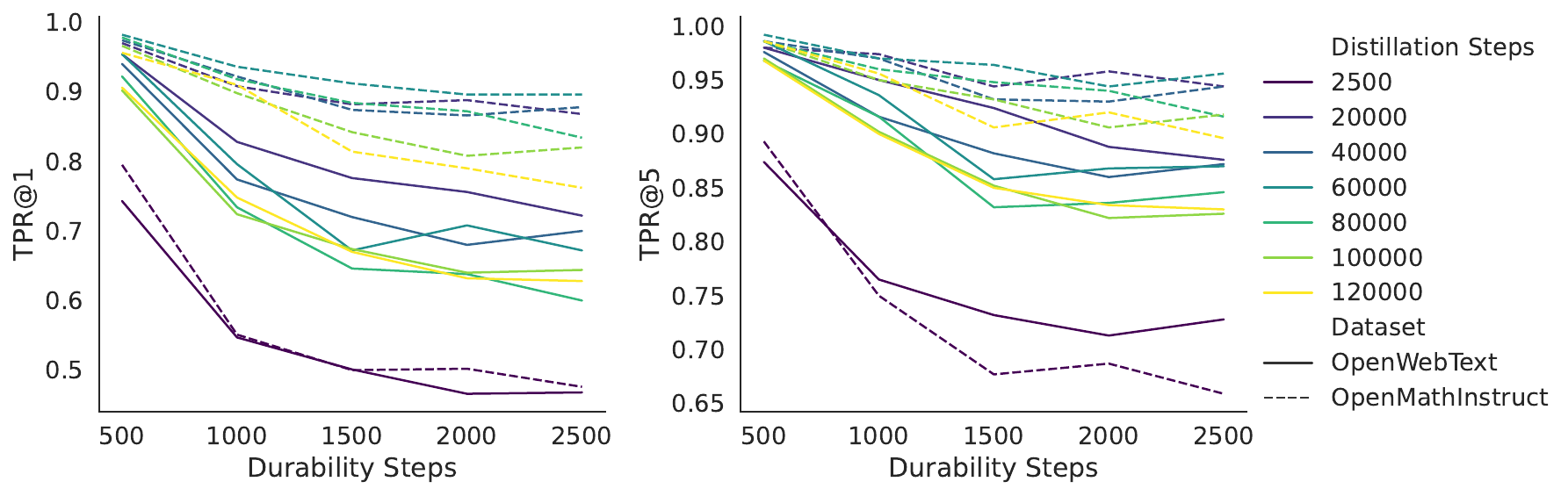}
    \caption{Evaluation of the TPR at 1\% and 5\% against finetuning for \KGW with increasing training dataset size when distilling the watermark on \textsc{GPT-2}.}
    \label{fig:ft_tpr_gpt2_ckpts}
\end{figure}

\fi

\clearpage

\end{document}